\documentclass[preprint,aps,nofootinbib,11pt,superscriptaddress]{revtex4-1}
\usepackage{geometry}                
\usepackage{amsmath,amssymb,amsfonts,dcolumn,color,graphicx,graphics,latexsym,placeins,epsfig,tikz}
\usetikzlibrary{arrows,shapes}
\usepackage{subfigure,rotating,bm,mathrsfs}
\usepackage[title]{appendix}
\usepackage{comment}

\newcommand{\be}{\begin{equation}}
\newcommand{\ee}{\end{equation}}
\newcommand{\ba}{\begin{eqnarray}}
\newcommand{\ea}{\end{eqnarray}}

\usepackage[pagebackref=false, colorlinks=true]{hyperref}
\definecolor{redish}{rgb}{0.7,0.2,0.0}  
\definecolor{bluish}{rgb}{0.2,0.5,0.8}

\hypersetup{linkcolor=redish,          
                  citecolor=blue,        
                  filecolor=magenta,      
                  urlcolor=blue}          

\begin{document}
\author{Kunal Pal}\email{kunalpal@iitk.ac.in}
\affiliation{Department of Physics, Indian Institute of Technology Kanpur, \\ Kanpur 208016, India}

\author{Kuntal Pal}\email{kuntal@iitk.ac.in}
\affiliation{Department of Physics, Indian Institute of Technology Kanpur, \\ Kanpur 208016, India}

\author{Rajibul Shaikh}\email{lrajibulsk@gmail.com}
\affiliation{Institute of Convergence Fundamental Studies,\\
School of Natural Sciences, College of Liberal Arts,\\
Seoul National University of Science and Technology, Seoul 01811, Korea}

\author{Tapobrata Sarkar}\email{tapo@iitk.ac.in}
\affiliation{Department of Physics, Indian Institute of Technology Kanpur, \\ Kanpur 208016, India}
\title{\Large A rotating modified JNW spacetime as a Kerr black hole mimicker}
\bigskip

\begin{abstract}
The Event Horizon Telescope has recently observed the images and shadows of the compact objects M87$^*$ and Sgr A$^*$ at the centres of the galaxies Messier 87 and Milky Way. 
This has opened up a new window in observational astronomy to probe and test gravity and fundamental physics in the strong-field regime. In this paper, we consider a rotating version of a 
modified Janis-Newman-Winicour metric, study its shadow, and constrain the metric parameters using the observed shadows of M87$^*$ and Sgr A$^*$. Depending on parameter values, 
the spacetime metric represents either a naked singularity or a wormhole. We find that the naked singularity case is not consistent with observations, as it casts a shadow which is 
much smaller than the observed ones. On the other hand, the shadow formed  by the  wormhole branch, depending on the parameter values, is consistent with the observations.
We put constraints on the wormhole throat radius by comparing the shadow with the observed ones of M87$^*$ and Sgr A$^*$.

\end{abstract}

\maketitle

\section{Introduction}
\label{intro}
 The remarkable results of the Event Horizon Telescope (EHT) collaboration, which captured the shadow of the ultra-compact supermassive objects at the centre of our and the nearby galaxy, has opened up a new era of observational astronomy \cite{EHT, EventHorizonTelescope:2019pgp,  EventHorizonTelescope:2022wkp, EventHorizonTelescope:2022xqj}. From the theoretical perspective, it is now possible to test the predictions of general relativity in very strong field regions with unprecedented accuracy. Even though the observed results in both the cases of M87$^*$ and Sgr A$^*$ are consistent with the canonical vacuum solution of general relativity, the Kerr metric \cite{carroll, EventHorizonTelescope:2022wkp}, other non-vacuum solutions of general relativity, or various modified gravity theories can not be completely ruled out yet.\footnote{ For recent works on constraining various parameters of these geometries, see \cite{Bambi}-\cite{shaikh:arxiv2022}.} One of the most important aspects of the study to this end is to look if various horizonless alternatives to black holes, like naked singularities and traversable wormholes are consistent with the EHT observations \cite{Cardoso:2019rvt, visser, RP1}. On the other hand, the presence of such exotic horizonless ultra-compact objects also motivates the study of regular models of black holes, where the singularity is replaced by a region of regular curvature \cite{Bardeen, Maeda, Lan:2023cvz}. Even though it is believed that the solution of a full quantum theory gravity will be regular, in the continued absence of a final theory of quantum gravity, various phenomenologically motivated approaches to resolve the singularity of a metric have gained significant attention in recent literature \cite{Maeda, Lan:2023cvz}.  
 
Among the various approaches indicated above, an elegant and promising one is that of Simpson and Visser (SV) \cite{SV1}, 
where the Schwarzschild singularity is replaced by a regular region, that can be timelike, null, or spacelike, depending 
on the parameter involved in the geometry. The final global geometry represents either a traversable wormhole or a regular 
black hole with one horizon. The importance of this approach is not only to resolve the Schwarzschild singularity but 
also to connect two different classes of spacetimes (black holes and wormholes) by a single parameter deformation (see \cite{SV2}-\cite{guo} 
for various applications and extensions of the SV method). An important direction to pursue here 
is to see whether the SV approach can be applied to that of a globally naked singularity, which was addressed recently in   \cite{Bronnikov, PPRS}. 
In particular, for the SV-modified version of the Janis-Newman-Winicour (JNW) spacetime\cite{JNW}, which is a naked singularity at a finite coordinate location, 
it was seen that the final metric can interpolate between a wormhole and a globally naked singularity 
\cite{PPRS}, which also provides a unifying version of two different class of metrics in the spirit of \cite{damour}. 

In this work, our main motivation is to test the viability of a rotating version of this class of spacetimes, and to constrain the relevant 
corresponding parameter space in light of the recent EHT data. To use the modified JNW metric as a realistic Kerr 
black hole mimicker, we have used the Newman-Janis algorithm \cite{Newman:1965my, Newman:1965tw} without 
complexification introduced by Azreg-A\"\i{}nou \cite{Azreg1, Azreg2, Azreg3}, to obtain an axially symmetric 
metric representing a rotating spacetime. We study the resulting phase structure in detail which reveals the presence of a 
wormhole as well as a naked singularity branch. Next, we discuss the separation of geodesics in this geometry 
and the consequent shadow formation. Finally, we will use the EHT data for M87$^*$ and Sgr A$^*$ to constrain 
the parameter space of our theory. 


\section{Rotating version of the modified JNW spacetime and its spacetime structure}

The  JNW metric \cite{JNW, Virbhadra, Virbhadra2} represents a static, spherically symmetric solution 
of the Einstein field equations in the presence of a minimally coupled scalar field, and is represented by the following line element,
\begin{equation}\label{JNW}
ds^2=-\Big(1-\frac{b}{r}\Big)^\gamma \text{d}t^2 + \Big(1-\frac{b}{r}\Big)^{-\gamma} \text{d}r^2 + 
\Big(1-\frac{b}{r}\Big)^{1-\gamma}r^2 \text{d}\Omega^2~.
\end{equation}
The parameters $b$ and $\gamma$  appearing above  are related to the ADM mass $M$ and the scalar charge $q$  through  the
following relations
\begin{equation}\label{parameters}
\gamma = \frac{2M}{b}~,~~b=2\sqrt{M^2+q^2}~.
\end{equation}
Note that $0\leq \gamma\leq 1$. The JNW metric reduces to the flat Minkowski spacetime and the Schwarzschild black hole for $\gamma=0$ ($M=0$) and $\gamma=1$ ($q=0$), 
respectively. For $0<\gamma<1$, it represents a globally naked singularity at the coordinate location $r=b$, as can be checked by calculating the scalar curvatures and 
from the analysis of null geodesics \cite{Virbhadra, Virbhadra2}. Throughout the paper, we focus on $0<\gamma<1$. 
In two recent works \cite{Bronnikov, PPRS}, the SV method of singularity resolution \cite{SV1} was applied to the JNW spacetime. In \cite{PPRS}, it was found that, 
depending on the interplay of the parameters, the singularity of the spacetime may or may not be resolved.\footnote{In \cite{Bronnikov}, the singularity was always resolved, due to the slight difference in choice of the coordinate system chosen to use the SV method.} The  form of the modified JNW (mJNW) metric obtained in \cite{PPRS} can be written as
\begin{equation}\label{eq:SVJNW}
ds^2=-\Big(1-\frac{b}{\sqrt{r^2+c^2}}\Big)^\gamma \text{d}t^2 + \Big(1-\frac{b}{\sqrt{r^2+c^2}}\Big)^{-\gamma} \text{d}r^2 
+ \Big(1-\frac{b}{\sqrt{r^2+c^2}}\Big)^{1-\gamma}\Big(r^2+c^2\Big) \text{d}\Omega^2~.
\end{equation}

Here $c$ (not the speed of light) is the SV parameter, which is a real and positive quantity having dimensions of length. This system represents a class of spacetimes that can interpolate between a two-way traversable wormhole and a globally naked singularity, hence is of importance in the context of modelling the geometry as a possible candidate for the supermassive ultra-compact at the centre of the Galaxies.

To study a more realistic situation, here we construct a rotating version of the mJNW metric by using a modification of the 
standard Newman-Janis algorithm that does not involve the complexification step and is proposed by Azreg-A\"\i{}nou \cite{Azreg1,Azreg2,Azreg3}. 
This procedure also guarantees that the final metric can be always written in Boyer-Lindquist (BL) 
type coordinates. It is important to note that the usual  Newman-Janis algorithm \cite{Newman:1965my, Newman:1965tw} 
that is successful in the case of the original 
SV metric to generate the rotating version \cite{shaikh:MNRAS2021, mazza,jcch,xutang}
can not be used here since the final metric can not be written in the BL coordinate. 

The systematic procedure to construct the rotating version is standard, and details can be found 
e.g. in \cite{Azreg1, Azreg2, Azreg3, Solanki}, and here we mention only the final form of the rotating metric. 
Starting  from a general static spherically symmetric metric of the form
\begin{equation}
ds^2=-G(r)\text{d}t^2+\frac{\text{d}r^2}{F(r)}+H(r)\text{d}\Omega^2~,
\label{eq:static_metric}
\end{equation}
the rotating version of this  metric after  the applying the  Azreg-A\"\i{}nou method can be written down as
\begin{eqnarray}
ds^2=-\frac{(FH+a^2\cos^2\theta)\psi}{(K+a^2\cos^2\theta)^2}\text{d}t^2+\frac{\psi}{FH+a^2}\text{d}r^2
-2a\sin^2\theta\left(\frac{K-FH}{(K+a^2\cos^2\theta)^2} \right)\psi \text{d}t \text{d}\phi+\psi \text{d}\theta^2 \nonumber \\ +\, \psi \sin^2\theta\left[1+a^2\sin^2\theta\left(\frac{2K-FH+a^2\cos^2\theta}{(K+a^2\cos^2\theta)^2} \right) \right]\text{d}\phi^2~.
\label{eq:rotating_aam}
\end{eqnarray}
Here, $K(r)=H\sqrt{\frac{F}{G}}$, and $\psi (r, \theta, a)=K(r)+a^2\cos^2\theta$, and $a$ is the specific angular momentum of the metric. The function
 $\psi (r, \theta, a)$ can be obtained as a solution of a complicated non-linear partial differential equation, 
 which we omit here, and can be found in \cite{Azreg1, Azreg2, Azreg3, Solanki}. In  the $a\rightarrow 0$ limit for the 
 above rotating metric to represent a normal fluid solution, it is required that $\lim_{a\to0} \psi(r, \theta, a)=H(r)$, 
 which implies that $F(r)=G(r)$.

With the mJNW metric in Eq.  \eqref{eq:SVJNW} as the starting point,  the final rotating metric in  BL coordinates can be rewritten as
\begin{equation}
ds^2 = -\left(1-\frac{2f}{\rho^2} \right)\text{d}t^2 + \frac{\rho^2}{\Delta} \text{d}r^2 + \rho^2 \text{d}\theta^2 
+ \frac{\Sigma\sin^2\theta}{\rho^2} \text{d}\phi^2 \nonumber \\ - \frac{4af\sin^2\theta}{\rho^2} \text{d}t \text{d}\phi ~,
\label{eq:SVJNW_rotating}
\end{equation}	
where 
\begin{eqnarray}
&& 2f (r) = (r^2+c^2) \left(1 - \frac{b}{\sqrt{r^2+c^2}} \right) \left[-1 + \left(1 - \frac{b}{\sqrt{r^2+c^2}} \right)^{-\gamma} \right]\,, \\
&& \rho^2 (r, \theta) = (r^2+c^2) \left(1 - \frac{b}{\sqrt{r^2+c^2}} \right)^{1-\gamma} + a^2\cos^2\theta\,, \\
&& \Delta (r)= (r^2+c^2) \left(1 - \frac{b}{\sqrt{r^2+c^2}} \right) + a^2\,, \\
&& \Sigma (r, \theta)= (\rho^2 + a^2\sin^2\theta)^2 - a^2\Delta\sin^2\theta~.
\end{eqnarray}
It is straightforward to check that this metric reduces to that of the rotating SV metric in the limit $\gamma \rightarrow 1$ \cite{mazza, shaikh:MNRAS2021} and that of the rotating version of the JNW metric in the limit $c \rightarrow 0$ \cite{Solanki}. Also, it can be checked that this metric reduces to that of the deformed JNW metric in the limit $a \rightarrow 0$. It is helpful to write the above metric in the following form 
\begin{equation}
ds^2=-\frac{\Delta}{\rho^2}(\text{d}t^2-a\sin^2\theta)^2+\frac{\rho^2}{\Delta}\text{d}r^2+\rho^2
\text{d}\theta^2+\frac{\sin^2\theta}{\rho^2}\big(a\text{d}t-(K+a^2)\text{d}\phi\Big)^2~,
\label{eq:rotating_canonical}
\end{equation}
where we have defined $K=(r^2+c^2) \left(1 - \frac{b}{\sqrt{r^2+c^2}} \right)^{1-\gamma}$.

To demystify the nature of the rotating metric constructed above, it is useful to make two successive 
coordinate transformations as in the non-rotating case \cite{PPRS}. These are given respectively by $\bar{r}^2=r^2+c^2$ and $\mathcal{R}^2=\Big(1-\frac{b}{\bar{r}}\Big)^{1-\gamma}\bar{r}^2$. 
After the first coordinate transformation, the radial component of the metric in Eq. \eqref{eq:rotating_canonical}
gets transformed to
\begin{equation}
g_{rr}\text{d}r^2=\frac{(r^2+c^2) \left(1 - \frac{b}{\sqrt{r^2+c^2}} \right)^{1-\gamma} 
		+ a^2\cos^2\theta}{(r^2+c^2) + a^2-b\sqrt{r^2+c^2}}\text{d}r^2= \frac{\bar{r}^2 \left(1 - \frac{b}{\bar{r}} \right)^{1-\gamma}+ a^2\cos^2\theta}{\bar{r}^2+a^2-b\bar{r}}\left(\frac{\bar{r}^2}{\bar{r}^2-c^2}\right)\text{d}\bar{r}^2~.
\end{equation}
Performing the second transformation mentioned above,  we can see that the final form of the radial part of  metric is given by
\begin{equation}
g_{\mathcal{R}\mathcal{R}}\text{d}\mathcal{R}^2=\left(\frac{\mathcal{R}^2+a^2\cos^{2}\theta}{\bar{r}^{2}(
	\mathcal{R})+a^2-b\bar{r}(\mathcal{R})}\right)
\frac{\bar{r}^{2}(R)}{\bar{r}^{2}(\mathcal{R})-c^2}
\frac{\text{d}\mathcal{R}^2}{\Big(1-\frac{b}{\bar{r}(\mathcal{R})}\Big)^{1-\gamma}
	\Big(1+\frac{1-\gamma}{2}\frac{b}{\bar{r}-b})\Big)^2}~.
\end{equation}
Here $\bar{r}$ implicitly depends on the radial coordinate $\mathcal{R}$ \cite{nandi1, nandi2}. We first observe that the radial 
part contains five singularities.
These are at locations
\begin{equation}
\bar{r}_{1}=c~,~~ \bar{r}_{2,3}=\frac{1}{2}\Big(b\pm\sqrt{b^2-4a^2}\Big)~, ~~\bar{r}_{4}=\frac{b(\gamma+1)}{2}~,~~\bar{r}_{5}=b~.
\end{equation}
In terms of the original radial coordinate $r$, these are given by
\begin{equation}
r_{1}=0~, ~r_{2,3}=\sqrt{\frac{1}{4}\Big(b\pm\sqrt{b^2-4a^2}\Big)^2-c^2}~, ~
r_{4}=\sqrt{\frac{b^2(\gamma+1)^2}{4}-c^2}~,~~\text{and}~~ r_{5}=\sqrt{b^2-c^2}~.
\end{equation} 
Note that, if it is real, then $r_{2,3}\leq r_5$ always. Therefore, we consider only $r_1$, $r_4$ and $r_5$ in our analysis.

Note that $0<\gamma<1$. Hence, the relevant singular point is $r_5$ when $b>c$, as $r_5>r_4>r_1$. Therefore, the coordinate $r$ (or $\bar{r}$) 
has the range $\sqrt{b^2-c^2}\leq r<\infty$ (or $b\leq \bar{r}<\infty$). A direct analysis of the Ricci curvature 
scalar reveals the presence of a 
curvature singularity at the location $r_{5}$ when $\theta=\pi/2$, which is similar to the ring singularity of a Kerr black hole. 
So, in this case, the metric represents a rotating naked singularity with the singularity being at $r_{5}=\sqrt{b^2-c^2}$ ($\bar{r}=b$).

On the other hand, when $b<c$, both $r_4$ and $r_5$ become imaginary, and the relevant singular point is $r_1=0$. 
However, one can check that the Ricci curvature scalar does not diverge at $r_1$, implying that this is a 
coordinate singularity. Since the $r$-dependence of the metric is only through the $r^2$ like terms, it can be 
extended to both the asymptotic infinities, $r\to \pm \infty$, through $r=0$, such that the metric can be thought of as 
two copies of the same spacetime glued together at $r=0$. Therefore, the metric in this case represents a 
two-way wormhole with the throat being at $r=r_1=0$ (or at $\bar{r}=\bar{r}_1=c$) \cite{morris}.

\section{Separation of the null geodesic equation}
We now describe the separation of the  null geodesic equation using the Hamilton-Jacobi (HJ) formalism. The  Hamilton-Jacobi equation  is given by
\begin{equation}\label{eq:Hamilton-Jacobi}
\frac{\partial S}{\partial \lambda}+\mathcal{H} = 0~, \;
\mathcal{H}=\frac{1}{2} g^{\mu \nu}p_{\mu}p_{\nu}~,
\end{equation}
with  $\lambda$ is the affine parameter, $S$ is the Jacobi action, and  $\mathcal{H}$ is the Hamiltonian. 
Also, $p_{\mu}$ is the four-momentum defined as
\begin{equation}
p_{\mu}=\frac{\partial S}{\partial x^{\mu}}=g_{\mu \nu}\frac{d x^{\nu}}{d\lambda}.
\label{eq:momentum_def}
\end{equation}
Since the metric, as well as the Hamiltonian in Eq. \eqref{eq:Hamilton-Jacobi} above do not depend explicitly on $t$ and 
$\phi$, we have two constants of motion -- the conserved energy $E=-p_t$ and the conserved angular 
momentum $L=p_\phi$ (about the axis of symmetry) \cite{Chandrasekhar}. Therefore, if there is any separable solution of 
Eq. \eqref{eq:Hamilton-Jacobi}, the Jacobi action can be written in terms of already known constants of the motion as 
\begin{equation}\label{eq:ansatz}
S = \frac{1}{2}\mu^2 \lambda - E t + L\phi + S_{r}(r) + S_{\theta}(\theta)~,
\end{equation}
where $\mu$ is the rest mass of the test particle, and $S_r(r)$, $S_\theta(\theta)$ are two functions of 
single variables, $r$ and $\theta$ respectively.
 For photons we have $\mu=0$. Substituting this  ansatz into the HJ Eq. \eqref{eq:Hamilton-Jacobi}, we obtain
\begin{equation}
\frac{1}{\psi}\Bigg[\left(FH+a^2\right)\left(\frac{dS_r}{dr}\right)^2-\frac{\Big(\left(K+a^2\right)E
	-aL\Big)^2}{FH+a^2}\Bigg]=\frac{1}{\psi}
\Bigg[-\left(\frac{dS_{\theta}}{d\theta}\right)^2-\frac{(aE\sin^2\theta-L)^2}{\sin^2\theta}\Bigg]~.
\end{equation}
Note that, for the case of motion in a null geodesic the function $\psi$ does not have any effect in separation, 
though this is not true for massive case. After some algebraic 
manipulations, we arrive at
\begin{equation}
\left(FH+a^2\right)\left(\frac{dS_r}{dr}\right)^2-\frac{\left(\left(K+a^2\right)E-aL\right)^2}{FH+a^2}
=-\Big(\frac{dS_{\theta}}{d\theta}\Big)^2-\frac{(aE\sin^2\theta-L)^2}{\sin^2\theta}~.
\end{equation}
The left-hand side of the above equation depends only on the radial coordinate $r$, and the right-hand side only depends on the angular coordinates $\theta$. Therefore, for the equality to hold, each should be equal to a constant,
which we denote as  $\mathcal{K}$. Writing $\mathcal{K}=C+(aE-L)^2$ for convenience, where $C$ is the Carter constant, we obtain
\begin{equation}
	(FH+a^2)^2\left(\frac{dS_r}{dr}\right)^2=\Big[E(K+a^2)-aL\Big]^2-(FH+a^2)\Big[C+(aE-L)^2\Big]~,~~\text{and}~~
\end{equation}
\begin{equation}
	\Big(\frac{dS_{\theta}}{d\theta}\Big)^2=C+a^2E^2\cos^2\theta-L^2\cot^2\theta ~.
\end{equation}
Finally, using Eq. \eqref{eq:momentum_def}, we obtain the following two separated  first order equations
corresponding to the null geodesic equation in the rotating spacetime constructed above 
\begin{equation}\label{psi}
	\psi\frac{dr}{d\lambda}=\pm E\sqrt{R(r)}~,\hspace{5mm}\text{and}~~~
    \psi\frac{d\theta}{d\lambda}=\pm E\sqrt{\Theta(\theta)}~,
\end{equation} 
where $R(r)$ and $\Theta(\theta)$ are given by
\begin{equation}\label{radial_potential}
 R(r) = \left[X(r) - a \xi \right]^2 - \Delta(r) \left[\eta + (\xi - a)^2 \right],
\end{equation}
\begin{equation}
	\Theta(\theta)=\eta+\cos^2\theta \Big(a^2-\frac{\xi^2}{\sin^2\theta}\Big)~.
\end{equation}
with  $\xi = \frac{L}{E}$ and $\eta = \frac{\mathcal{K}}{E^2}$.
We have also used the notation
\begin{equation}
	X(r)=K+a^2~,\hspace{5mm}\Delta(r)=F(r)H(r)+a^2~,
\end{equation}
where both are functions of the radial coordinate $r$ only.
It is important to note that both $R(r)$ and $\Theta(\theta)$ are  greater than or equal to zero. 
The equations for the $t$ and $\phi$ coordinates are omitted, as those are not required to study the shadow structure. Note that we get the separated geodesic equations 
in the rotating mJNW when we use the appropriate form of $F(r)$, $G(r)$ and $H(r)$ in the above equations.

\section{SHADOW structure OF THE ROTATING METRIC}

For a black hole, the contour of the shadow is formed by unstable circular orbits of photons. However, as we discuss below, the situation is a bit different for naked singularities and wormholes. For convenience, we write down the radial geodesic equation
(given by the first equation in \eqref{psi} above)  in terms of $\bar{r}$ coordinate.
This is given by 
\begin{equation}
\psi\frac{d\bar{r}}{Ed\lambda}=\pm \sqrt{\bar{\Delta}(\bar{r})} \sqrt{R(\bar{r})}~,
\end{equation}
where
\begin{equation}
R(\bar{r})=\left[X(\bar{r}) - a \xi \right]^2 - \Delta(\bar{r}) \left[\eta + (\xi - a)^2 \right],
\end{equation}
\begin{equation}
\bar{\Delta}(\bar{r})= 1-\frac{r_{0}^2}{\bar{r}^2}, \quad \Delta(\bar{r})=\bar{r}^2-r_s\bar{r}+a^2, ~~\text{and}~~
X(\bar{r})=\bar{r}^2\left(1-\frac{r_s}{\bar{r}}\right)^{1-\gamma}+a^2.
\end{equation}
Here, $r_0=\bar{r}_1=c$ is the throat radius in the wormhole case, and $r_s=\bar{r}_5=b$ is the naked singularity location in the naked singularity case. The unstable photon orbits are given by the conditions
\begin{equation}
\dot{\bar{r}}=0~, \quad \ddot{\bar{r}}=0~, \quad \dddot{\bar{r}} \geq 0~,
\label{eq:unstable_condition}
\end{equation}
where a dot represents differentiation with respect to $\lambda$. We now discuss the naked singularity and the wormhole cases separately.

\subsection{The naked singularity case}

For the naked singularity case, $\bar{\Delta}(\bar{r})$ is always positive and never vanishes in the whole range of $\bar{r}$. Therefore, the unstable photon orbit 
conditions in Eq. \eqref{eq:unstable_condition} become
\begin{equation}\label{eq:photon_sphere}
R(\bar{r}_{ph})=0~, \quad R'(\bar{r}_{ph})=0~, \quad R''(\bar{r}_{ph}) \geq 0~,
\end{equation} 
where $\bar{r}=\bar{r}_{ph}$ is the unstable photon orbit radius. The first two conditions give \cite{Shaikh:2019fpu}
\begin{equation}\label{eq:xi}
\xi=\left[\frac{X(\bar{r})}{a}-\frac{2\Delta(\bar{r})X'(\bar{r})}{a\Delta'(\bar{r})}\right]_{{\bar{r}=\bar{r}_{ph}}}~,
\end{equation}
\begin{equation}\label{eq:eta}
\eta= \frac{4a^2 (X'(\bar{r}))^2\Delta(\bar{r})-\big[(X(\bar{r})-a^2)
\Delta'(\bar{r})-2X'(\bar{r})\Delta(\bar{r})\big]^2}{a^2(\Delta'(\bar{r}))^2}\bigg|_{\bar{r}=\bar{r}_{ph}}~,
\end{equation}
where a prime denotes a derivative w.r.t to $\bar{r}$. $\xi$ and $\eta$ in the above equations give the critical 
impact parameters of the unstable photon orbits forming the contour of the shadow.

To find the apparent shape of the shadow as seen by an observer, we need to use the so called celestial coordinates $\alpha$ and $\beta$ defined as
\begin{equation}\label{eq:celestial1}
\alpha=\lim_{\bar{r}_o\to\infty}\left(-\bar{r}_o^2\sin\theta_o\frac{d\phi}{d\bar{r}}\Big\vert_{(\bar{r}_o,\theta_o)}\right),
\end{equation}
\begin{equation}\label{eq:celestial2}
\beta=\lim_{\bar{r}_o\to\infty}\left(\bar{r}_o^2\frac{d\theta}{d\bar{r}}\Big\vert_{(\bar{r}_o,\theta_o)}\right),
\end{equation}
where $(\bar{r}_o,\theta_o)$ is the position of a faraway observer. After using separated geodesic equations, we obtain
\begin{equation}\label{eq:alpha}
\alpha=-\frac{\xi}{\sin\theta_o}~,
\end{equation}
\begin{equation}\label{eq:beta}
\beta=\pm \sqrt{\eta+a^2\cos^2\theta_o-\xi^2\cot^2\theta_o}~.
\end{equation}
The contour of the shadow is constructed by using $\bar{r}_{ph}$ as a parameter and then plotting parametric plots of $\alpha$ and $\beta$ using Eqs. \eqref{eq:xi}, \eqref{eq:eta}, \eqref{eq:alpha} and \eqref{eq:beta}.

For a black hole, all the photons whose impact parameters lie inside the shadow contour plunge inside the unstable 
photon orbits and get captured by the event horizon, thereby creating a shadow. For a naked singularity also,
 we must check whether or not all those photons which plunge inside the unstable orbits get captured by the singularity. 
 To this end, we first find out $R(\bar{r})$ at $\bar{r}=r_s$. This is given by
\begin{equation}
R(r_s)=a^2(a^2\cos^2\theta_o-\alpha^2\cos^2\theta_o-\beta^2)~, \quad  ~~\text{for} ~~~0<\gamma<1~,
\end{equation}
where we have used Eqs. \eqref{eq:alpha} and \eqref{eq:beta} to replace $\xi$ and $\eta$ by $\alpha$ and $\beta$. The impact parameters $\xi$ and $\eta$ (or $\alpha$ and $\beta$) of all the photons having turning point just at the singularity satisfy $R(r_s)=0$, which in turn gives, for $a\neq 0$,
\begin{equation}
\frac{\alpha^2}{a^2}+\frac{\beta^2}{a^2\cos^2\theta_o}=1~.
\label{eq:shadow_NS}
\end{equation}
This is an ellipse with semi-major axis $a$ and semi-minor axis $a\cos\theta_o$. In the region outside
the ellipse, $R(r_s)<0$, and therefore ingoing photons with $\alpha$ and $\beta$ lying in this outside region get turned away from some outer 
radius $\bar{r}_{tp}$ where $R(\bar{r}_{tp})=0$ and do not get absorbed. Therefore, this outside region cannot be a part of the shadow. On the other hand, $R(r_s)>0$ in the inside region 
of the ellipse. Also, we find that for $\alpha$ and $\beta$ lying in this inside region, $R(\bar{r})>0$ for all
 $\bar{r}>r_s$. Therefore, ingoing photons with $\alpha$ and $\beta$ lying in the inside region get absorbed. We  
 also notice that the curve in Eq. \eqref{eq:shadow_NS} completely lies inside the $\alpha-\beta$ curve given by unstable circular orbits, i.e., by the curve (if it exists) obtained from Eqs. \eqref{eq:xi}, \eqref{eq:eta}, \eqref{eq:alpha} and \eqref{eq:beta}. This implies that, unlike black hole cases, in this naked singularity case, photons which plunge inside the unstable circular orbits get turned away from outside the naked singularity if their $\alpha$ and $\beta$ lie in between the two curves just mentioned and get absorbed if their $\alpha$ and $\beta$ lie inside the ellipse \eqref{eq:shadow_NS}. Therefore, in this naked singularity case, the shadow is given only by the ellipse in Eq. \eqref{eq:shadow_NS}.

\subsection{The wormhole case}
In this case, for unstable circular orbits lying outside the wormhole throat $r_0$, the conditions in Eq. \eqref{eq:unstable_condition} turns out to be $R(\bar{r}_{ph})=0$, $R'(\bar{r}_{ph})=0$ and $R''(\bar{r}_{ph}) \geq 0$. 
The first two conditions give Eqs. \eqref{eq:xi} and \eqref{eq:eta}. Therefore, the part of the shadow formed by 
these unstable circular orbits is given by the curve obtained from Eqs. \eqref{eq:xi}, \eqref{eq:eta}, \eqref{eq:alpha} and \eqref{eq:beta}. However, it is also known that the wormhole throat can also act as a location of unstable 
circular orbits \cite{shaikh1,shaikh2,shaikh:MNRAS2021,shaikh:arxiv2022, shaikh3}. For such orbits, $\bar{\Delta}(\bar{r}_{ph})=0$, and $\dot{\bar{r}}(\bar{r}_{ph})=0$ is satisfied automatically as $\bar{r}_{ph}=r_0$. Therefore, for such orbits, 
the conditions in Eq. \eqref{eq:unstable_condition} now turn out to be $R(r_0)=0$, $R'(r_0)>0$. 
The first condition gives
\begin{equation}
\left[X(r_0)+a\sin\theta_o\alpha\right]^2-\Delta(r_0)\left[\beta^2+\left(\alpha+a\sin\theta_o\right)^2\right]=0~,
\label{eq:shadow_WH}
\end{equation}
where we have used Eqs. \eqref{eq:alpha} and \eqref{eq:beta} to replace $\xi$ and $\eta$ by $\alpha$ and $\beta$. 
The above equation gives the part of the shadow contour which is formed by the unstable photon orbits located at the throat. 
The complete shadow is given by the common region enclosed by the curve \eqref{eq:shadow_WH} and the curve obtained from Eqs. \eqref{eq:xi}, \eqref{eq:eta}, \eqref{eq:alpha} and \eqref{eq:beta}. For more detailed explanation of a shadow in 
a wormhole case, see \cite{shaikh:MNRAS2021,shaikh:arxiv2022}.

\begin{figure}[h]
\centering
\subfigure{\includegraphics[scale=0.6]{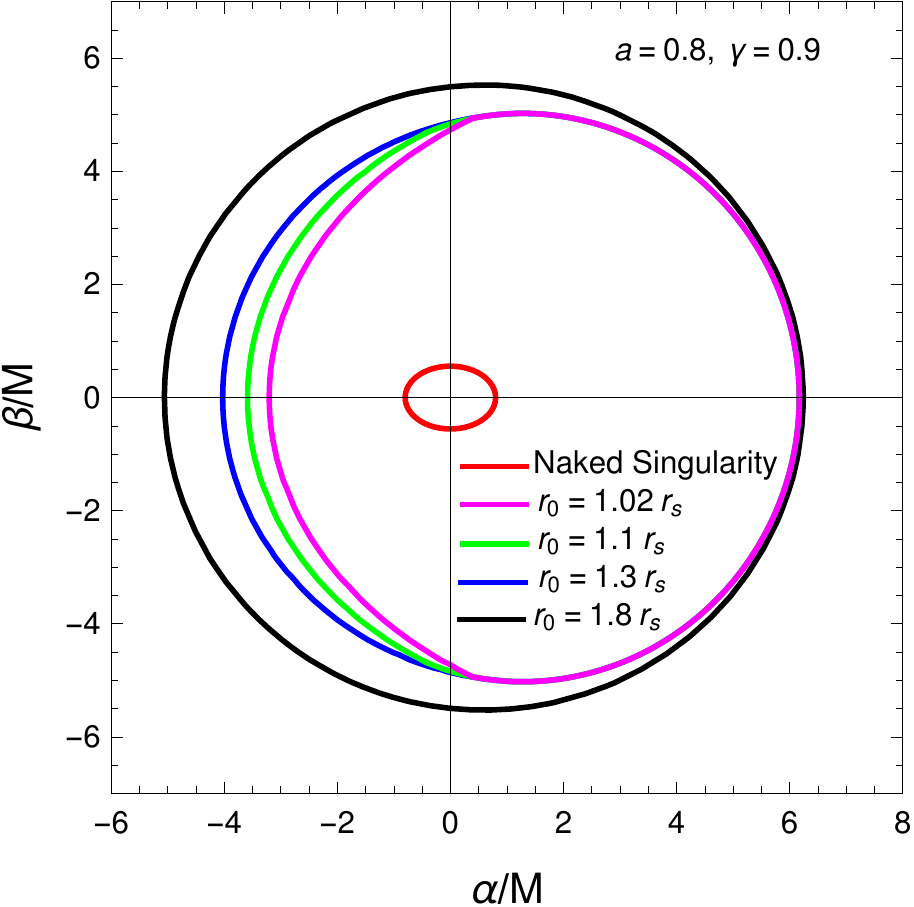}}
\subfigure{\includegraphics[scale=0.6]{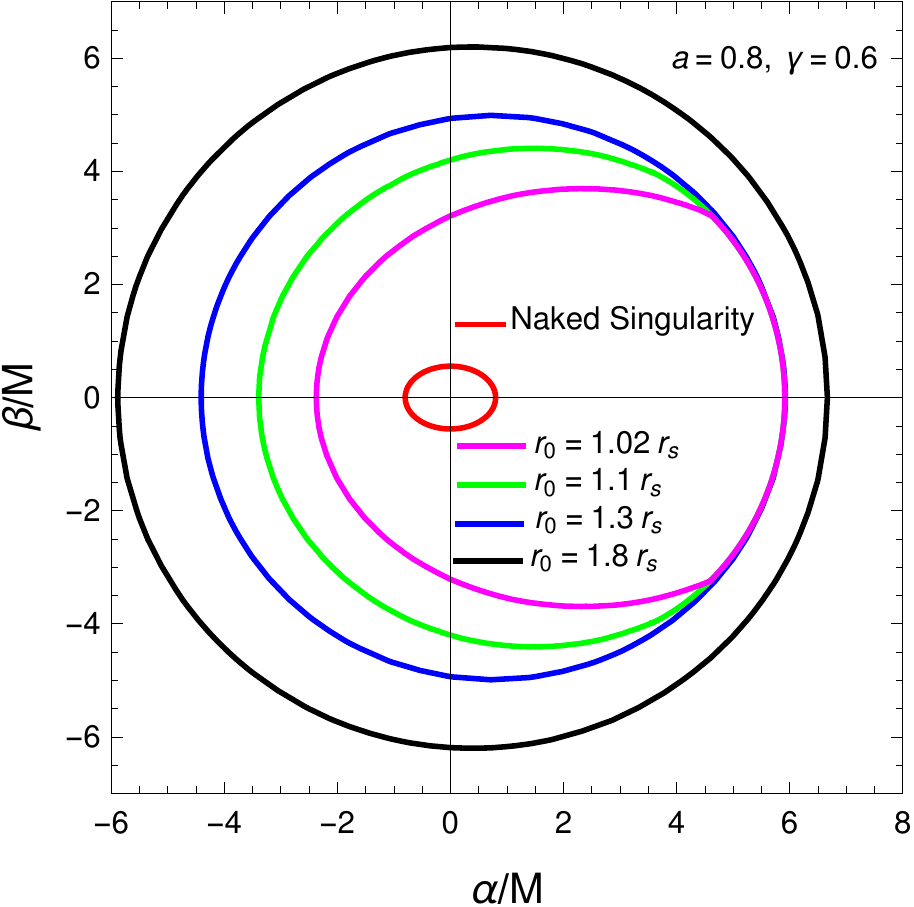}}
\subfigure{\includegraphics[scale=0.6]{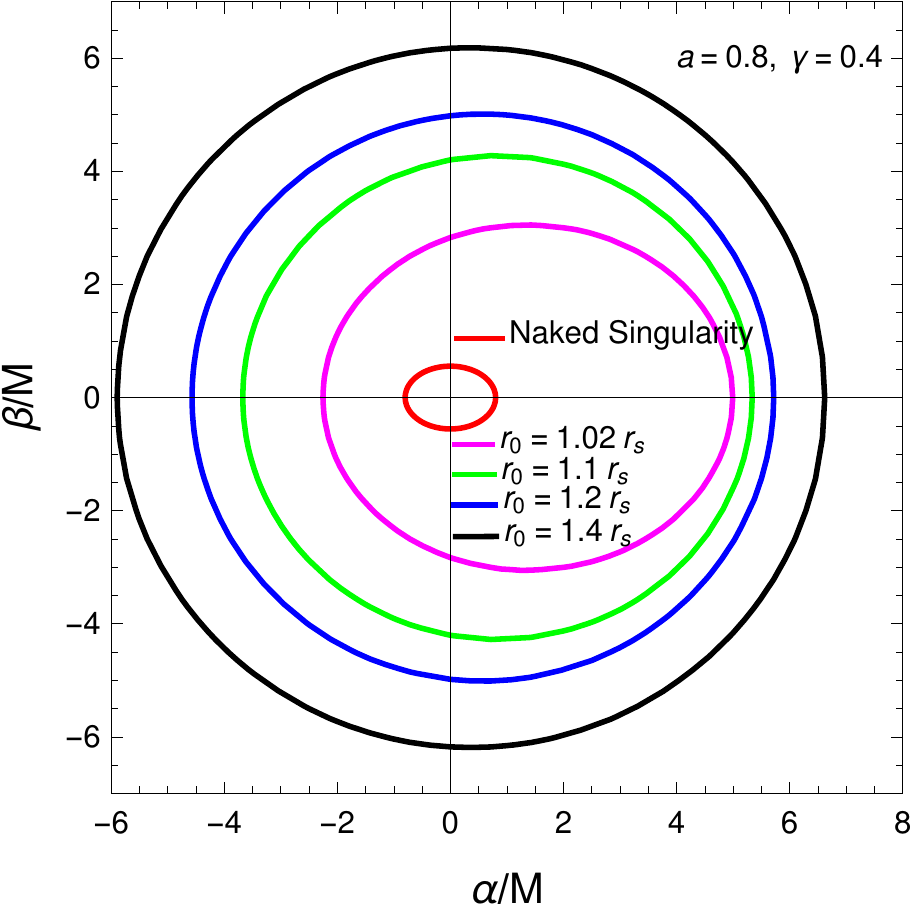}}
\caption{Shadows cast by the rotating mJNW metric for the inclination angle $\theta_o=46^\circ$. The red curve is the 
	shadow contour of the naked singularity case. All other cases are due to the wormhole case, and the corresponding
	 parameters are provided in the figure.}
\label{fig:MSRV_shadow}
\end{figure}

Figure \ref{fig:MSRV_shadow} shows some characteristics shadow structures  of the 
rotating mJNW metric. Note that the shadow in the naked singularity case is very small. As we shall see in the next 
section, this case is not consistent with M87$^*$ and Sgr A$^*$ observations. In the wormhole case, for a given 
spin $a/M$ and the inclination angle $\theta_o$, the shadow size increases and becomes more and more circular with 
increasing throat size $r_0$.

\section{Constraining the metric from the M87$^*$ and Sgr A$^*$ observations}

In this section, we check the viability of the rotating mJNW metric as a candidate for the super massive object at the 
centre of the two galaxies reported by the EHT. To this end, we use the EHT results of the recent observation of the shadows of 
M87$^*$ and Sgr A$^*$ and put possible constraints on the parameters of the rotating  geometry constructed above.

For our purpose, we use the average size of the shadow and its deformation from circularity. The shadow contour 
is perfectly circular for zero spin, but with the increasing value of the spin, it starts deforming from the 
circular shape, and also, its center in the $\alpha-\beta$ plane starts shifting from the origin. However, as the 
shadow has a reflection symmetry about the $\alpha$-axis, the corresponding geometric center $(\alpha_c,\beta_c)$ 
is given by $\alpha_c=(1/A)\int \alpha dA$ and $\beta_c=0$, $dA$ being an area element. Therefore, the average 
radius $R_{sh}$ of the shadow is given by \cite{Bambi}
\begin{equation}
R_{sh}^2=\frac{1}{2\pi}\int_{0}^{2\pi}l^2(\phi)\; d\phi~,
\end{equation}
where $l(\phi)=\sqrt{(\alpha(\phi)-\alpha_c)^2+\beta(\phi)^2}$, and $\phi=\arctan(\beta(\phi)/(\alpha(\phi)-\alpha_c))$ 
is the angle between the $\alpha$-axis and the vector connecting the geometric centre $(\alpha_c,\beta_c)$ with a point $(\alpha,\beta)$ on the shadow boundary. Following \cite{EHT}, we define the deviation $\Delta C$ from circularity as
\begin{equation}
\Delta C=\frac{1}{R_{sh}}\sqrt{\frac{1}{2\pi}\int_{0}^{2\pi}(l(\phi)-R_{sh})^2\; d\phi}~.
\end{equation}
The angular diameter of the shadow is given by $\Delta\theta_{sh}=2R_{sh}/D$, where $D$ is the distance to the center of the galaxy.

As we have seen in the previous section, the shadow of the naked singularity is given by the ellipse in Eq. \eqref{eq:shadow_NS}. Therefore, in this case, the geometric center and the average radius, respectively, turns out to be $(\alpha_c,\beta_c)=(0,0)$, and $R_{sh}=a\sqrt{\cos\theta_o}$.

\subsection{Constraints from the M87$^*$ observation:}

As reported by the EHT collaboration for the M87$^*$ observation, we will use $D=(16.8\pm 0.8)$ Mpc and the mass of the object to be $M=(6.5\pm 0.7)\times 10^9 M_\odot$ \cite{EHT}. The inclination angle is taken to be $\theta_o=17^\circ$. 
According to the EHT collaboration, the angular size of the observed shadow is $\Delta\theta_{sh}=42\pm 3$ $\mu$as \cite{EHT} 
and the spin lies within the range $0.5\leq a/M\leq 0.94$. Also, the deviation from circularity is reportedly less than 
$10\%$, i.e., $\Delta C\leq 0.1$. Therefore, the observed value of the shadow diameter in dimensionless units is 
estimated to be
\begin{equation}
\frac{d_{sh}}{M}=\frac{D\Delta\theta_{sh}}{M}=11.0\pm 1.5~,
\end{equation}
where the errors have been added in quadrature.

In case of the naked singularity, the calculated dimensionless shadow diameter for the inclination angle $\theta_o=17^\circ$ turns out to be $d_{sh}/M=2a\sqrt{\cos\theta_o}/M\simeq 1.96\;a/M$. Note that, for the spin range given above, this is much smaller than the observed value. We therefore conclude that the naked singularity case is not consistent with observations.

\begin{figure}[h]
\centering
\subfigure[]{\includegraphics[scale=0.57]{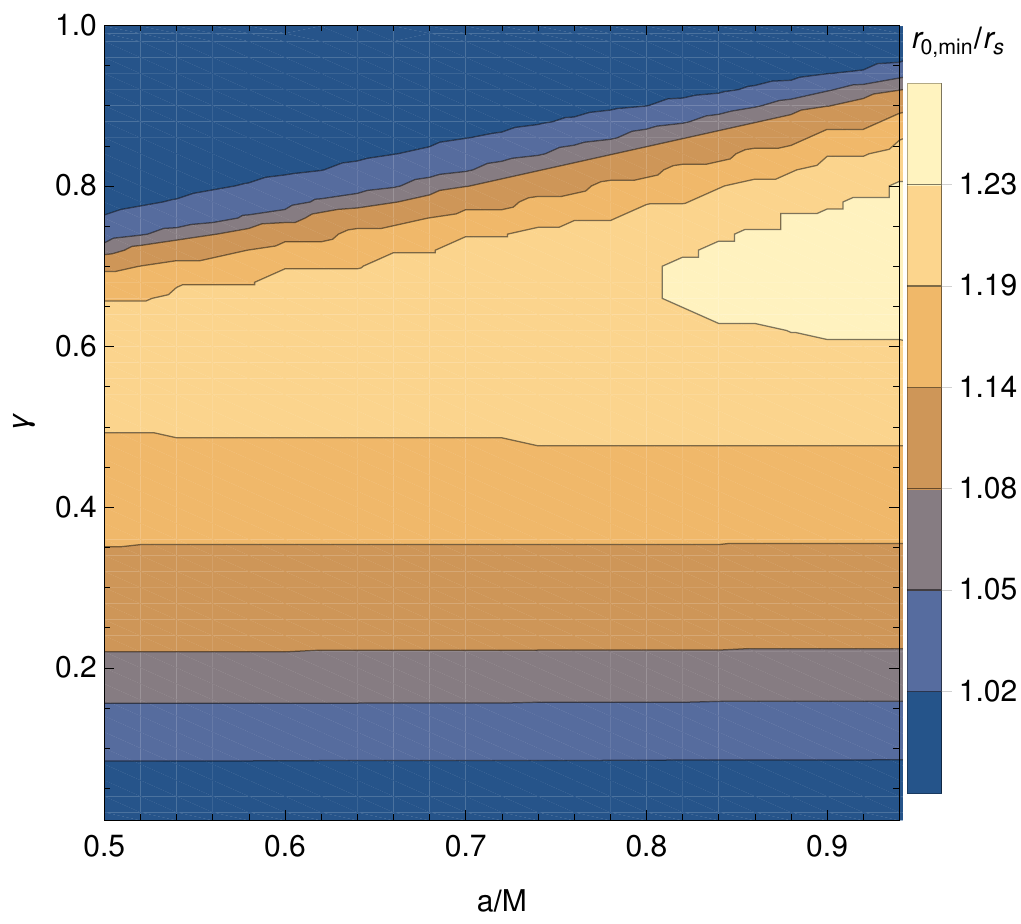}}
\subfigure[]{\includegraphics[scale=0.57]{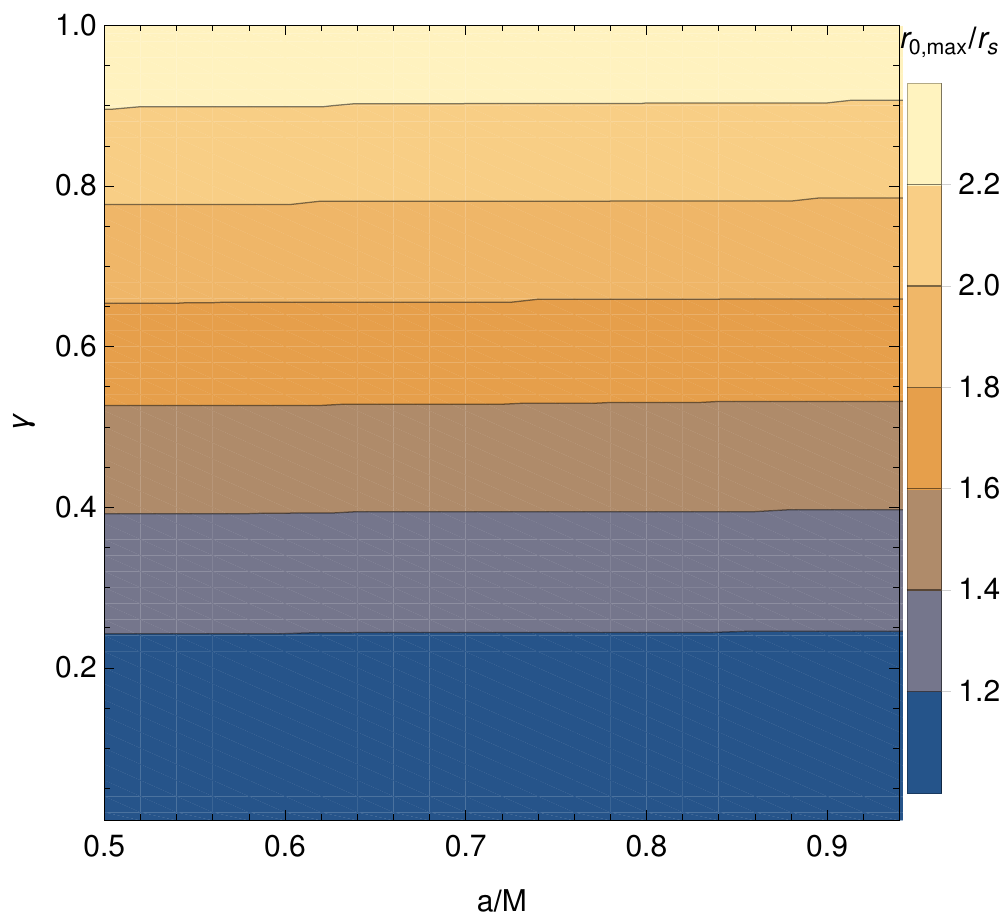}}
\subfigure[]{\includegraphics[scale=0.57]{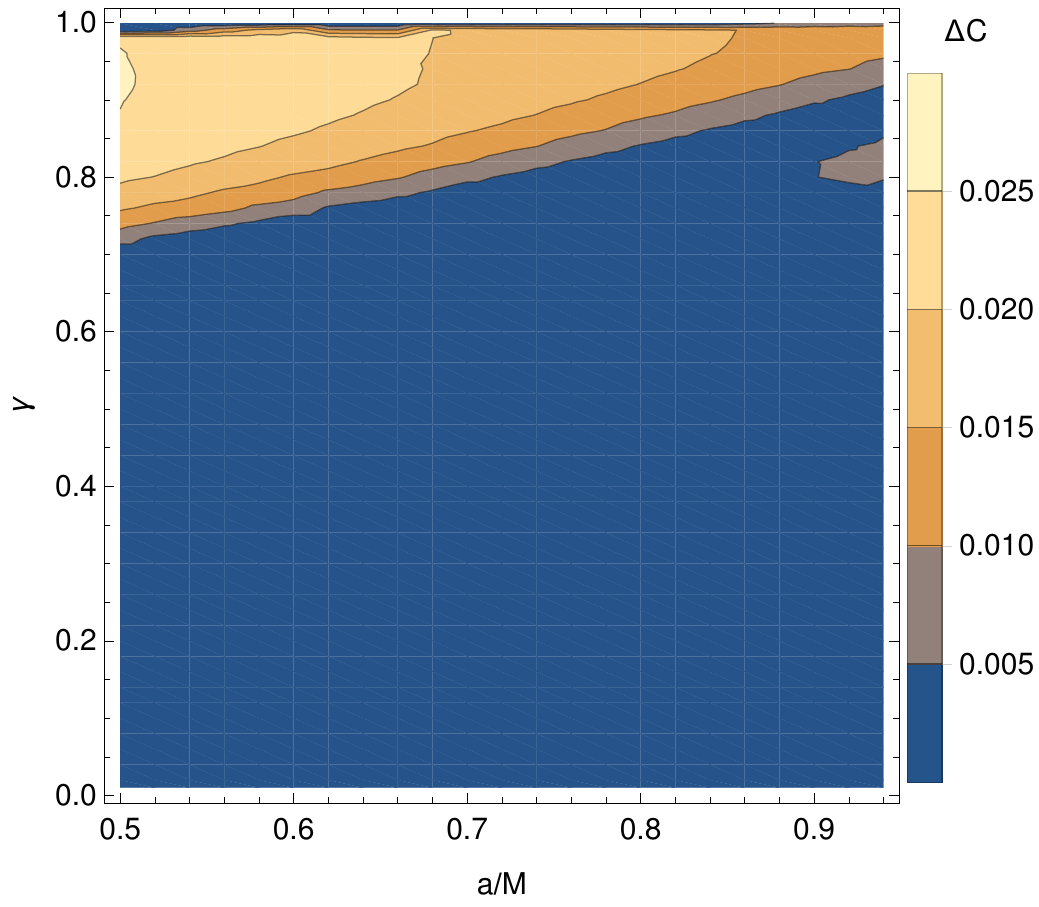}}
\caption{Plots showing the spin $a/M$ and $\gamma$ dependence of (a) the minimum, (b) the maximum throat radius allowed by the M87$^*$ observation and (c) the maximum deviation from circularity. The inclination angle is taken to be $\theta_o=17^\circ$.}
\label{fig:M87_constraining}
\end{figure}

As we have seen for the wormhole case in the previous section, for a given spin $a/M$ and the inclination angle $\theta_o$, the shadow size 
increases and becomes more and more circular with increasing throat size $r_0$. We therefore calculate the minimum $r_{0,min}$ 
and the maximum $r_{0,max}$ throat radius for which the shadow size corresponds to the allowed lower value $9.5$ and upper 
value $12.5$, respectively. The maximum $\Delta C$ in the range $r_{0,min}\leq r_0\leq r_{0,max}$ will be at $r_0=r_{0,min}$. 
In Fig. \ref{fig:M87_constraining}, $r_{0,min}$, $r_{0,max}$ and maximum $\Delta C$ are shown. 
Note that maximum $\Delta C$ is much lesser than the allowed maximum value $0.1$. Therefore, for a given $\gamma$ 
and the spin $a/M$, the shadow in the wormhole case is consistent with the M87$^*$ observation if 
$r_{0,min}\leq r_0\leq r_{0,max}$. 

\subsection{Constraints from the Sgr A$^*$ observation:}
The recent papers on Sgr A$^*$ observations have used the fractional deviation parameter $\delta$ to constrain different spacetimes \cite{EventHorizonTelescope:2022xqj}. Here, $\delta$ is a measure of the fractional deviation of the shadow diameter from that of a Schwarzschild black hole and is given by
\begin{equation}
\delta=\frac{d_{sh}}{d_{sh,Sch}}-1=\frac{R_{sh}}{3\sqrt{3}M}-1=\frac{\hat{R}_{sh}}{3\sqrt{3}}-1~,
\end{equation}
where $\hat{R}_{sh}=R_{sh}/M$ is the dimensionless radius of the shadow and $d_{sh,Sch}=6\sqrt{3}M$ is the diameter of the Schwarzschild black hole shadow in units where the 
Newton's gravitational constant and the speed of light are set to unity. 
Using the observed angular diameter of the Sgr A$^*$ shadow and two separate sets of information, such as the distance $D$ and mass $M$ of Sgr A$^*$ from the Very Large Telescope Interferometer (VLTI) and Keck observations, the EHT collaboration has provided the following bounds on the fractional deviation parameter $\delta$ \cite{EventHorizonTelescope:2022wkp, EventHorizonTelescope:2022xqj}
\begin{equation}
\delta=\left\{
\begin{array}{ll}
-0.08^{+0.09}_{-0.09}  & \;\;(\mbox{VLTI}) \\
-0.04^{+0.09}_{-0.10}  & \;\;(\mbox{Keck})
\end{array}
\right..
\end{equation}
Therefore, the fractional deviation parameter lies in the range $-0.17\leq \delta\leq 0.01$ (VLTI) and $-0.14\leq \delta\leq 0.05$ (Keck). In the following, we will use these bounds to constrain the rotating metrics under consideration. Although the EHT collaboration has not provided any constraint on the spin and $\Delta C$, according to them, inclination more than $50^\circ$, i.e., $\theta_o\geq 50^\circ$ is disfavoured. Here we use $\theta_o=134^\circ$ (or equivalently $46^\circ$) as provided by the GRAVITY Collaboration \cite{GRAVITY}.

In the naked singularity case, the calculated $\delta$ for $\theta_o=46^\circ$ turns out to be $\delta=[a\sqrt{\cos\theta_0}/(3\sqrt{3})-1]\simeq (0.16a/M-1)$ which, for the spin range $0<a/M<1$, lies outside both the VLTI and Keck bound. Therefore, the naked singularity branch of the rotating metric constructed above 
 is not consistent with the Sgr A$^*$ observation as well.

\begin{figure}[h]
\centering
\subfigure{\includegraphics[scale=0.65]{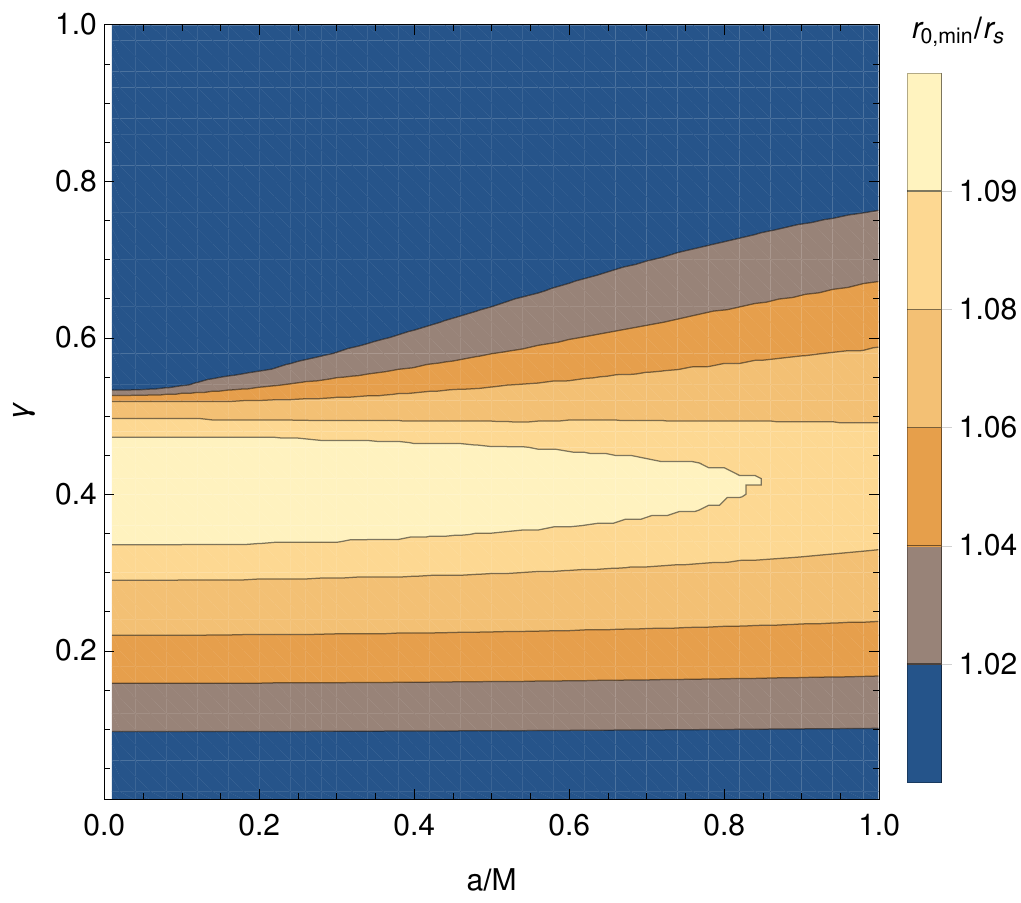}}
\subfigure{\includegraphics[scale=0.65]{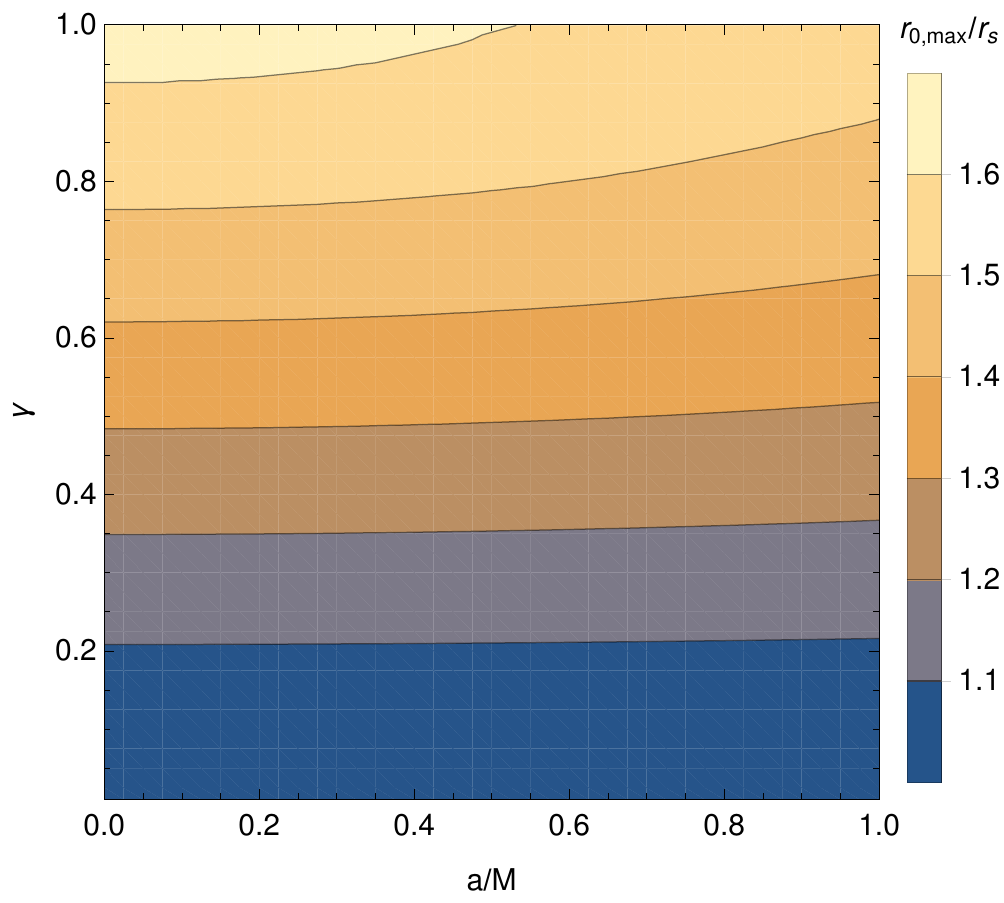}}
\caption{Plots showing the spin $a/M$ and $\gamma$ dependence of the minimum $r_{0,min}$ and the maximum throat radius $r_{0,max}$ allowed by VLTI bound. The inclination angle is taken to be $\theta_o=46^\circ$.}
\label{fig:Sgr_constraining_vlti}
\end{figure}

\begin{figure}[h]
\centering
\subfigure{\includegraphics[scale=0.65]{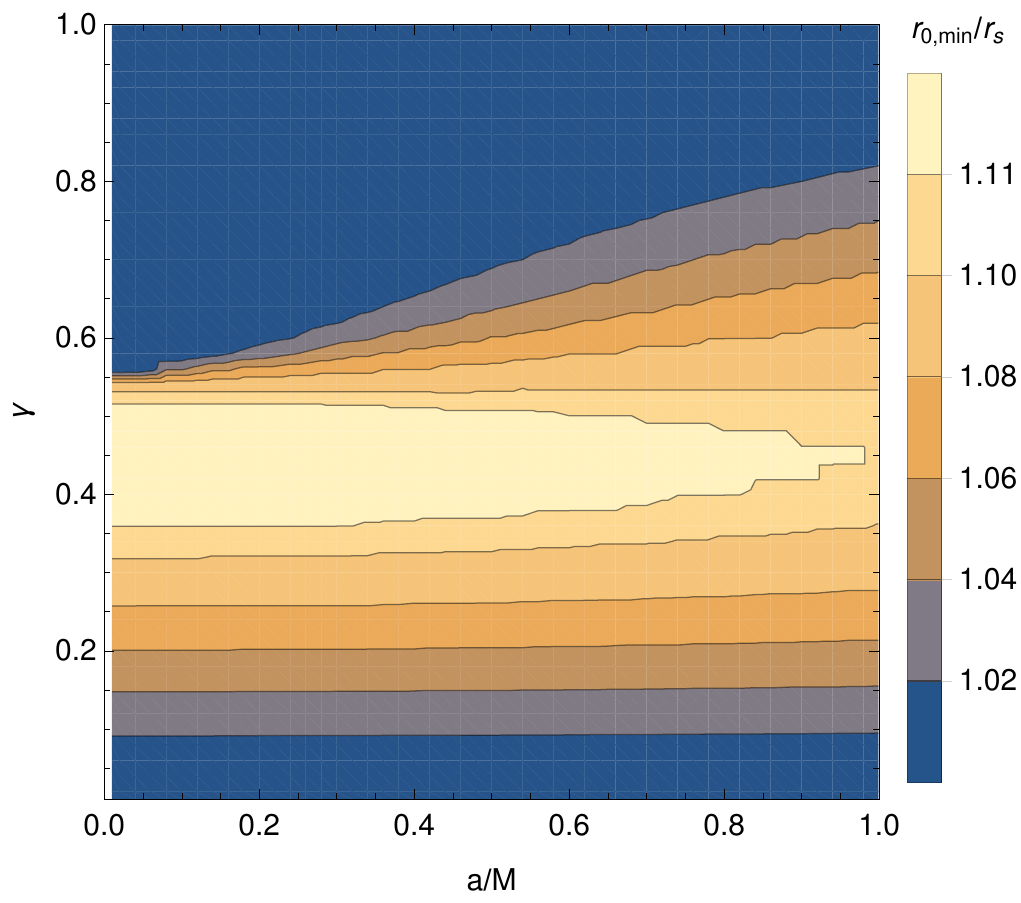}}
\subfigure{\includegraphics[scale=0.65]{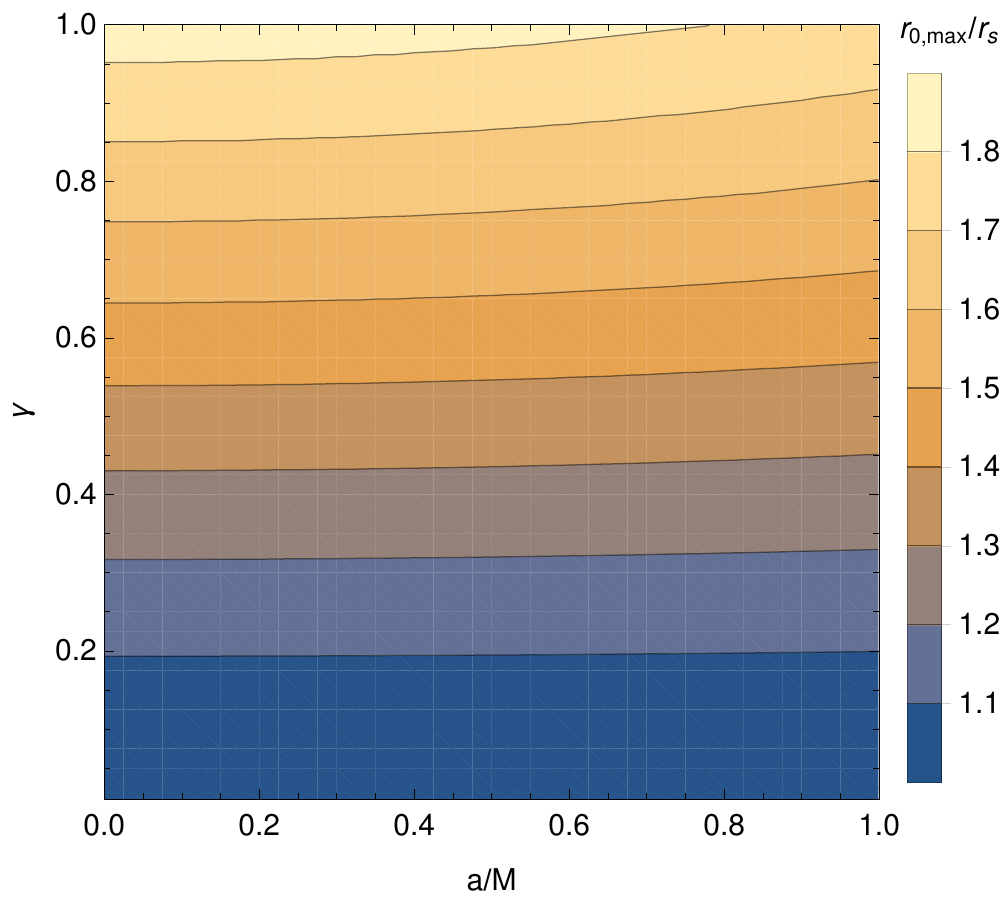}}
\caption{Plots showing the spin $a/M$ and $\gamma$ dependence of the minimum $r_{0,min}$ and the maximum throat radius $r_{0,max}$ allowed by Keck bound. The inclination angle is taken to be $\theta_o=46^\circ$.}
\label{fig:Sgr_constraining_keck}
\end{figure}

For the wormhole case, following the same method as in the previous subsection, we calculate the minimum $r_{0,min}$ and the maximum $r_{0,max}$ throat radius for which $\delta$ corresponds to the allowed lower and upper value, respectively. They are shown in Figs. \ref{fig:Sgr_constraining_vlti} and \ref{fig:Sgr_constraining_keck}. Therefore, for a given $\gamma$ and the spin $a/M$, the shadow in the wormhole case is consistent with the Sgr A$^*$ observation if $r_{0,min}\leq r_0\leq r_{0,max}$.

\section{Conclusions}
In this paper, we have constructed a rotating version of the mJNW metric, studied its shadow structure, and constrained the metric using the EHT observations of both M87$^*$ and Sgr A$^*$. The mJNW metric represents either 
a naked singularity or a wormhole. We have found that the naked singularity is not consistent with the observations as 
it casts a shadow that is much smaller than the observed ones. On the other hand, the wormhole casts a shadow that, 
depending on the parameter values, is consistent with the observations. We have found out the lower bound $r_{0,min}$ and the upper bound $r_{0,max}$ on the throat radius $r_0$. These bounds depend on the metric parameter $\gamma$ and the spin $a/M$. The shadow in the wormhole case is consistent with the observed shadows of M87$^*$ and Sgr A$^*$ if the throat radius lies in between these bounds, i.e., if $r_{0,min}\leq r_0\leq r_{0,max}$. Tighter constraint on the throat radius can be put if the exact spin of M87$^*$ and Sgr A$^*$ and deformation from the circularity of the observed shadow of Sgr A$^*$ are known in the future.

\section*{Acknowledgements}
The work of RS is supported by the grant from the National Research Foundation funded by the Korean government, No. NRF-2020R1A2C1013266.

{}

\end{document}